# Laser Pulse Driven Terahertz Generation via Resonant Transition Radiation in Inhomogeneous Plasmas

Chenlong Miao[a], John P. Palastro[b] and Thomas M. Antonsen[a]

[a]Insititute for Research in Electronics and Applied Physics, University of Maryland, College Park 20742

[b]Naval Research Laboratory, Washington DC 20375

An intense, short laser pulse propagating across a plasma boundary ponderomotively drives THz radiation. Full format PIC simulations and theoretical analysis are conducted to investigate the properties of this radiation. Simulation results show the THz emission originates in regions of varying density and covers a broad spectrum with maximum frequency close to the maximum plasma frequency. In the case of a sharp vacuum-plasma boundary, the radiation is generated symmetrically at the plasma entrance and exit, and its properties are independent of plasma density when the density exceeds a characteristic value determined by the product of the plasma frequency and the laser pulse duration. For a diffuse vacuum-plasma boundary, the emission from the plasma entrance and exit is asymmetric: increasing and decreasing density ramps enhance and diminish the radiated energy respectively. Enhancements by factors of 50 are found and simulations show that a 1.66 J, 50 fs driver pulse can generate ~400 µJ of THz radiation in a 1.2 mm increasing density ramp. We present a model that attributes this effect to a plasma resonance process in the density ramp. The results from the model match those of the simulations for ramp lengths less than 600 µm. For longer ramps for which simulations are too time consuming the model shows that the amount of radiation reaches a maximum at a ramp length determined by collisional absorption.

# I. INTRODUCTION

Electromagnetic terahertz radiation (THz) spans frequencies from 300 GHz to 20 THz and has a wide variety of applications[1] including spectroscopy[2], remote detection[3], and medical and biological imaging[4]. Intense THz pulses can be generated at large scale accelerator facilities via synchrotron or transition radiation, but the size and cost of such facilities can be prohibitive for widespread use. Laser-solid interactions provide a small-scale alternative, but material damage limits the THz radiation to µJs per pulse[5]. Plasmas, on the other hand, can withstand large amplitude optical pulses, motivating laser-plasma interactions as a path to high efficiency small-scale THz sources[6-7].

There are a number of schemes in which THz radiation is generated via laser plasma interactions. A first example is the transition radiation generated by a laser accelerated electron bunch passing from plasma to vacuum[8-12]. In an experiment reported in Ref. 8, coherent radiation in the range of 0.3-3 THz was produced by this mechanism using nC, femtosecond electron bunches. The THz energy collected within a limited 30 mrad angle was ~3-5 nJ and was observed to increase quadratically with the bunch charge. A second example is the radiation produced during two-color laser pulses gas ionization[13-15]. A fundamental and second harmonic optical field combined with correct relative phase produce a slow, directed, photoionization current that can drive THz radiation. This mechanism has been shown to routinely produce peak THz energy in excess of 5 µJ with pump pulse energy of tens of mJ. A third example is the coherent radiation involving a laser pulse propagating through a plasma channel with an axially periodic density profile[16-17]. This slow wave structure supports electromagnetic channel modes with subluminal phase velocities and further allows phase matching between the laser driver and modes, which provides the possibility of high conversion efficiency. Simulation shows THz energy of 6 mJ is generated with a pulse energy of 0.5 J with a depletion length of less than 20 cm.

Here, using theory and simulation, we investigate ponderomotively driven THz radiation that occurs as a laser pulse crosses a plasma boundary[18]. We will refer to this mechanism as transition radiation in analogy with the charged particle beam counterpart. The resulting THz radiation is characterized by conical emission and a broad spectrum with maximum frequency occurring near the maximum plasma frequency[19]. We find that the amount of THz radiation is substantially enhanced when the laser pulse passes through a gradual increasing density ramp. In particular, a 1.66 J, 50 fs driver pulse can generate ~400 µJ of THz radiation in a 1.2 mm increasing density ramp, comparing quite favorably with existing THz generation schemes.

The organization of this paper is as follows. In Section II, we introduce the transition radiation mechanism in the case a laser pulse normally incident on a sharp plasma boundary. A differential equation describing the radiation excitation in 2D planar geometry with an arbitrary density profile is presented. This equation can be solved for a sharp boundary, providing an integral expression for the radiated spectrum. Section III presents analysis and simulations of THz generation when a laser pulse is incident on a diffuse plasma boundary. An increasing density ramp results in resonant transition radiation, which enhances the radiation relative to the sharp plasma boundary. We derive a scaling law describing how the radiation depends on the plasma and laser parameters. We also show that substantially less radiation is generated in a decreasing density ramp. In Sec. IV we present our conclusions and discuss future directions.

## II. THz generation via Transition Radiation in a sharp vacuum-plasma interface

We begin by demonstrating the THz generation mechanism using the full format PIC simulation TurboWAVE[20]. The simulations, conducted in the lab frame, consist of a finite sized target plasma illuminated by a laser pulse incident from the left. Fig. 1a displays an example for the case of a sharp boundary, uniform plasma. A range of laser pulse and plasma parameters is considered. To quantify the radiation emitted from the plasma, we calculate the Poynting flux through prescribed surfaces outside the plasma region. This captures the radiation emitted in the forward, backward and lateral directions.

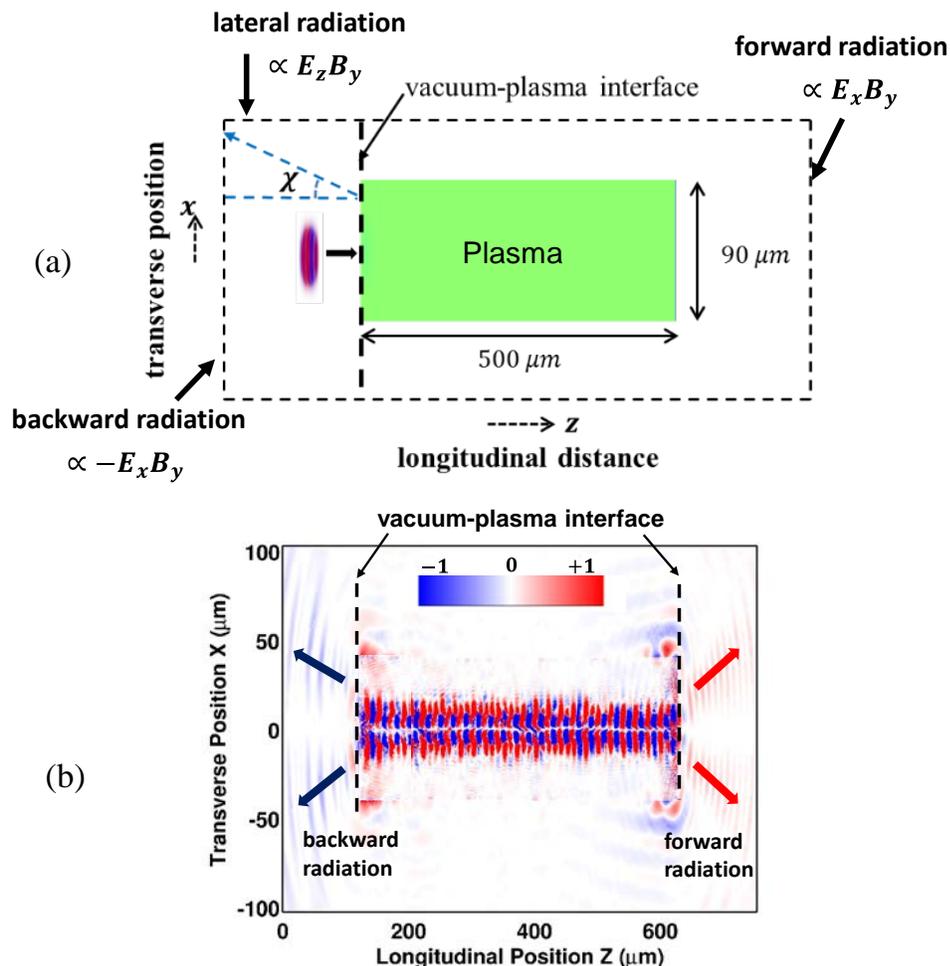

Figure 1: (a) Diagram of simulation set up for transition radiation at the vacuum plasma boundary. The diagnostic box is set outside the plasma channel. The Poynting flux through each surface is calculated. (b) A snap shot of the Poynting flux $E_x B_y$ in PIC simulations using TurboWAVE[20]. The transition radiation is generated at both boundaries when the laser travels through the plasma from left to right.

As an example, we consider a uniform plasma with sharp step boundaries as illustrated in Fig. 1a. A laser pulse of duration (FWHM) $\tau_P = 50$ fs traverses the plasma from left to right. The plasma is 500 µm long and 90 µm wide with a density of $n = 2.8 \times 10^{18}$ cm$^{-3}$. The transverse plasma dimension is chosen mainly for the convenience of simulation. Fig. 1b shows a false color image of the longitudinal component of the Poynting flux after the laser pulse traverses the entire plasma from left to right. Plasma wave excitation can be observed as the rapid longitudinal oscillations in the Poynting flux. The alternate positive and negative values of Poynting flux indicate a small average flux. However, at both ends outside the plasma, we observe transition radiation as the red and blue streaks denoting forward and backward Poynting flux, respectively.

The laser pulse produces a low frequency ponderomotive force, which is proportional to the gradient of the laser pulse intensity, $F_P \sim -\nabla I$. This force drives the ponderomotive currents that produce the THz radiation. We note that a key element that is required for coherent ponderomotively driven THz generation is that the plasma currents be dynamic in all inertial reference frames. This occurs in Refs. 16 and 17 by virtue of the fact that the driving laser pulse propagates in a spatially periodic plasma channel. It occurs in Refs. 18, 19 and the present work due to the fact that the ponderomotively driven transition radiation is generated at the boundary of the plasma. Thus, transition radiation is a possible explanation for the results of the Refs. 6 and 7. However, THz radiation will not be generated by the currents in the steady wake following a laser pulse in a uniform plasma as suggested by the analysis of Refs. 6 and 7. In such a case, all currents are steady in a frame moving with the group velocity of the laser pulse, and consequently no radiation occurs.

Following Ref. 18, we can derive an expression for the radiated energy per unit frequency ($\omega$) and per unit angle in 2D planar geometry. We take the laser pulse intensity and hence ponderomotive potential to be described by a spatio-temporal Gaussian profile,

$$V_p = V_{p0} \exp\left(-\frac{x^2}{R_L^2} - \frac{\xi^2}{L_p^2}\right), \tag{1}$$

where $V_{p0} = m_e c^2 a_0^2 / 4$ is the maximum of the laser ponderomotive potential in terms of the normalized laser vector potential $a_0 = eE_0 / (m_e c \omega_0)$ with the laser central frequency $\omega_0$ and field amplitude $E_0$. The quantities $\sqrt{\ln 2} R_L$ and $2\sqrt{\ln 2} L_p$ are the FWHM laser spot size and pulse duration respectively. The quantity $L_p = c\tau_p$ is the length and $\xi = z - ct$ is the shifted longitudinal coordinate.

As detailed in Appendix A, we obtain the following differential equation for the Fourier transformed transverse component of the electric field $\bar{E}_x(k_x, z, \omega)$, which is driven by the ponderomotive potential of the laser pulse

$$\frac{d}{dz}\left(\frac{\varepsilon(z)}{k^2(z)} \frac{d}{dz} \bar{E}_x\right) + \varepsilon(z) \bar{E}_x \equiv S(z) = -i \frac{d}{dz}\left(\frac{k_x}{k^2(z)} \frac{\omega_p^2(z)}{\omega^2} \frac{d}{dz} \frac{\bar{V}_p}{q_e}\right) - i k_x \frac{\omega_p^2(z)}{\omega^2} \frac{\bar{V}_p}{q_e},$$

(2)

where $\varepsilon(z) = 1 - \omega_p^2(z)/\omega(\omega + i\nu)$ is the frequency dependent dielectric function of the plasma and $\nu$ is the collision frequency, which in our case is much less than the radiation frequency $\omega$. In obtaining Eq. (2) we have assumed the electron density depends only on the longitudinal direction $z$. That is, the plasma has infinite transverse extent. The longitudinal wavenumber is defined as $k^2 = \omega^2 \varepsilon / c^2 - k_x^2$ and $k_z = \omega / c$. The quantity $\bar{V}_p = \hat{V}_p(\omega, k_x) \exp(ik_z z)$ is the Fourier transformed amplitude of the laser pulse $V_p$ with $\hat{V}_p$ given by the following expression:

$$\hat{V}_p = V_{p0} \pi R_L \tau_p \exp\left(-\frac{\omega^2 \tau_p^2}{4} - \frac{k_x^2 R_L^2}{4}\right),$$

(3)

The radiated energy per unit length $U'$ can be obtained in the form of a spectral density in $(\omega, k_x)$

$$\frac{dU'}{d\omega} = \int_{-\infty}^{\infty} dk_x S_z(\omega, k_x) \tag{4a}$$

$$U' = \int_0^{\infty} d\omega \frac{dU'}{d\omega}. \tag{4b}$$

The spectral density $S_z$ is given by

$$S_z(\omega, k_x) = \frac{c}{8\pi^2} \frac{1}{2\pi} \left( E_x(\omega, k_x) B_y^*(\omega, k_x) + c.c. \right), \quad (5)$$

where $B_y$ is the transverse component of the magnetic field, which in turn is given by[18]

$$B_y = i \frac{\omega}{c} \frac{1}{k^2(z)} \left( k_z k_x \frac{\omega_p^2(z)}{\omega^2} \frac{V_p}{q_e} - \varepsilon(z) \frac{dE_x}{dz} \right). \quad (6)$$

In 2D planar geometry the nonzero field components are $(E_x, B_y, E_z)$ and in 2D cylindrical geometry the field components are $(E_r, B_\theta, E_z)$, thus the radiated THz will be polarized with electric field $E$ in the r-z plane.

In Appendix A, Eq. (2) is solved in the sharp boundary approximation for the plasma density $n_e = 0$ for $z < 0$ (vacuum) and $n_e = n_0$ for $z \geq 0$. The resulting radiated energy per unit frequency ($\omega$) and length radiated backward in 2D planar geometry is expressed as,

$$\frac{dU'}{d\omega} = \frac{L_p^2 R_L^2}{|\omega|} \frac{\omega_{p0}^4}{32\pi} \frac{m_e^2}{q_e^2} a_0^4 \int_0^1 d\alpha \frac{\alpha^2 \sqrt{1-\alpha^2} \times \exp\left[ -\frac{\omega^2}{2c^2} \left( L_p^2 + R_L^2(1-\alpha^2) \right) \right]}{\left| \varepsilon(\omega)\alpha + \sqrt{\alpha^2 - \omega_{p0}^2/\omega^2} \right|^2}. \quad (7)$$

The integration over $\alpha$ can be viewed as an integration over all possible transverse wavenumbers $k_x$ since $\alpha = \cos\chi = \sqrt{1 - k_x^2 c^2 / \omega^2}$, where $\chi$ is observation angle as shown in Fig. 1a. Thus the integrand in Eq. (7) approximates the angular distribution of the radiation. We note that the integrand vanishes for both $\alpha = 0$ and $\alpha = 1$. Therefore the emerging radiation will have a conical distribution. The simulation using the same parameters as in Fig. 2 shows that the radiation peeks at an angle $\chi = 25°$.

In Fig. 2 backward radiation spectra predicted by the simulation are compared with Eq. (7). For this comparison, the laser pulse energy is 66mJ ($a_0 = 0.4$) with a 800 nm wavelength, 15 μm spot size and 50 fs pulse duration. Below the maximum plasma frequency, a broad spectrum of radiation is observed. Results are shown for two different length plasma slabs demonstrating that, as expected, the radiation is insensitive to the plasma length. This is consistent with the radiation originating from the left boundary. Comparing the theory and simulation one notices a large fluctuation around $\omega/\omega_p \approx 0.4$ in the simulation results labeled L = 500 µm and L = 100 µm, which is not observed in the 2D theory. This fluctuation is dependent on the transverse size of the target plasma. In particular, if we increase the transverse size from 90 µm to 180 µm, we got the curve labelled W = 180 µm on Fig. 2 in which the fluctuation is partially suppressed. We speculate that the fluctuation is due to the interference between ray paths involving reflection from the side boundary. One may also notice a low frequency cut-off ($0.15\omega/\omega_p$) for the simulation due to the finite size of simulation domain.

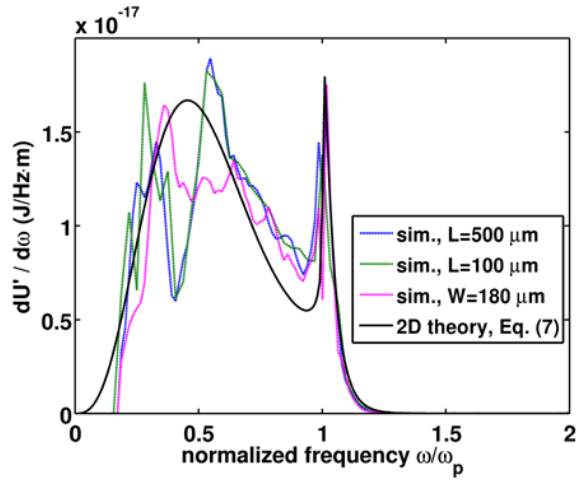

Figure 2: Comparison of radiated spectral density. Shown are theoretical values (Eq. (7), black solid), and simulations using TuroWAVE for plasma length L= 500 µm (blue dashed) and L = 100 µm (green dashed). Simulation result for plasma with a larger transverse size W = 180 µm (magenta dashed) is also provided.

Figure 3 displays the dependence of the total energy per unit length $U'$ on plasma density. The radiated energy is nearly constant for densities above $1.5 \times 10^{18} cm^{-3}$. In the inset Fig. 3(b), the spectra of radiated energy for different densities is displayed, along with the frequency spectrum of the ponderomotive potential. The spectra of the ponderomotive potential limits the radiated energy spectrum, explaining the saturation of the radiated energy with density. For densities above the characteristic density determined by the laser pulse duration, the laser pulse excites a current at the boundary that depends primarily on the properties of the laser pulse. For example in Fig. 3, the peak radiated energy occurs at an electron density of $1.5 \times 10^{18}\,\text{cm}^{-3}$, and we have $\omega_{p0}\tau_p \sim \pi$.

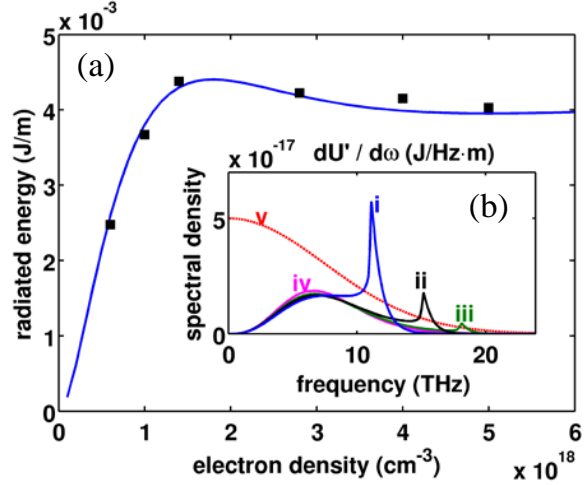

Figure 3: (a) Radiated energy versus electron density (2D theory: line; TurboWAVE: squares.). Energy is insensitive to plasma density above $1.5 \times 10^{18}\,\text{cm}^{-3}$. (b) Radiation spectra for different plasma densities: (i) $1.5 \times 10^{18}\,\text{cm}^{-3}$, (ii) $2.8 \times 10^{18}\,\text{cm}^{-3}$, (iii) $4 \times 10^{18}\,\text{cm}^{-3}$, (iv) $6 \times 10^{18}\,\text{cm}^{-3}$. The red dotted line (v) is the spectrum of the laser envelope $\overline{V}_p$ with arbitrary units.

We have also conducted simulations to examine the scaling of radiated energy with laser intensity. According to Eq. (7), the radiated energy should scale as $a_0^4$ when $a_0 \ll 1$. The simulations verified this scaling. However, as $a_0$ increases into the relativistic regime, the radiation saturates possibly due to the increase in effective electron mass resulting from the quiver velocity. As an example, the radiated energy is about 140 µJ with $a_0 = 4$ and 48 nJ with $a_0 = 0.4$.

# III. THz generation via Resonant Transition Radiation in diffuse plasma profiles

The results in Sec. II are for the case of a sharp plasma-vacuum. When the density transition has a ramp the results change in two ways. First there is an asymmetry between radiation generated when the laser enters and leaves the plasma. Second, the amount of radiation generated as a pulse enters the plasma increases as the length of the transition region increases. The asymmetry is shown in Fig. 4, where false color images of the spectrum of the Poynting flux through the lateral simulation boundary are displayed. For a sharp boundary (4a), as discussed in Sec. II, the radiation generated by the laser pulse entering and leaving the plasma is the same as evidenced by the equal size patches in (4b). However, when the ramp is added (4c) the amount of energy radiated when the laser enters the plasma goes up while the amount radiated upon leaving goes down (4d).

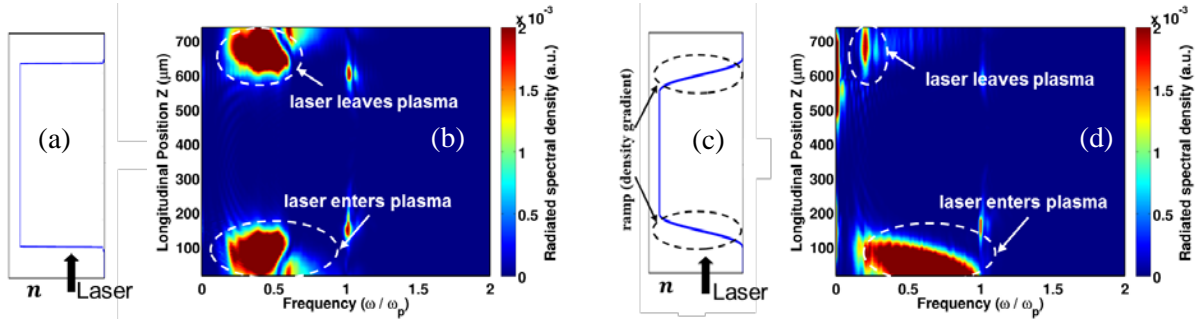

Figure 4: Comparisons of (a) plasma density with sharp step boundaries and (c) plasma density with increasing and decreasing ramps (~25 µm). Simulation results, radiation spectrum as a function of the longitudinal distance and frequency showing that (b) the step boundary is symmetric while (d) the case with density ramps shows asymmetry.

In Fig. 5 we plot the total energy per unit length radiated from the increasing density transition for several ramp lengths. Also shown in this figure are the results obtained by solving Eq. (2) and evaluating Eq. (4a). The solid line in Fig. 5 is a scaling formula that will be derived subsequently.

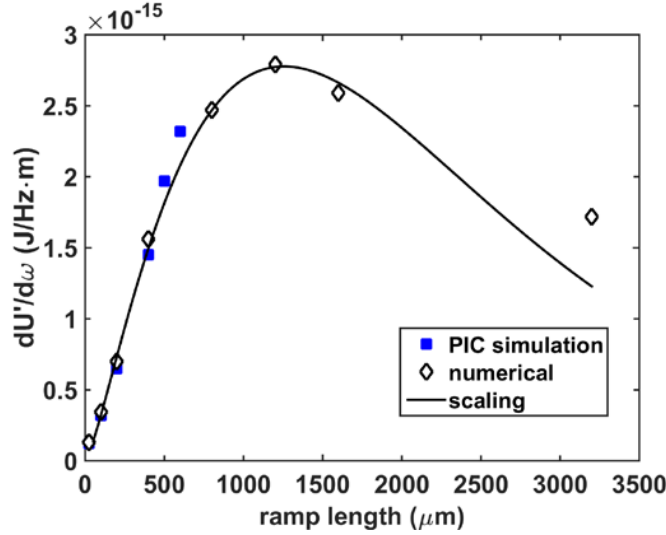

Figure 5: Comparisons of radiated spectral density using different increasing ramp lengths at a particular radiation frequency $\omega_1 = 0.6\omega_{p0}$. Shown are simulation results (square), numerical results of our model (diamond) and the scaling theory Eq. (11).

To investigate the effect of a density ramp we turn to numerical solutions of Eq. (2) for the THz electric field in the case of diffuse density profiles. Shown in Fig. 6 are the simulation results of the electric field $\bar{E}_x(\omega, k_x)$ at a particular radiation frequency $\omega = 0.8\omega_{p0}$ and wavenumber $k_x = 0.4\omega/c$ where $\omega_{p0}$ is the maximum plasma frequency and $c$ is the speed of light in vacuum. The laser properties and electron density remains the same as in Sec. II. Figs. 6a and 6b show that the laser pulse propagates through an increasing density ramp ( $k_z > 0$ )a decreasing ramp, respectively. This is realized in the solution of Eq. (2) not by reflecting the density profile, but rather, by changing the direction of the laser propagation. This in turn is done by making the substitution $\bar{V}_p = \hat{V}_p \exp(ik_z z)$ to $\bar{V}_p = \hat{V}_p \exp(-ik_z z)$ in Eq. (2). In both cases, Fig. 6a and 6b, the field in the uniform density region ( $z > 200$ μm ) responds to the laser pulse in the form of a wake with a spatial dependence in the form of a wave, $\exp(\pm ik_z z)$. This is the particular solution given by Eq. (B1) of Appendix B.

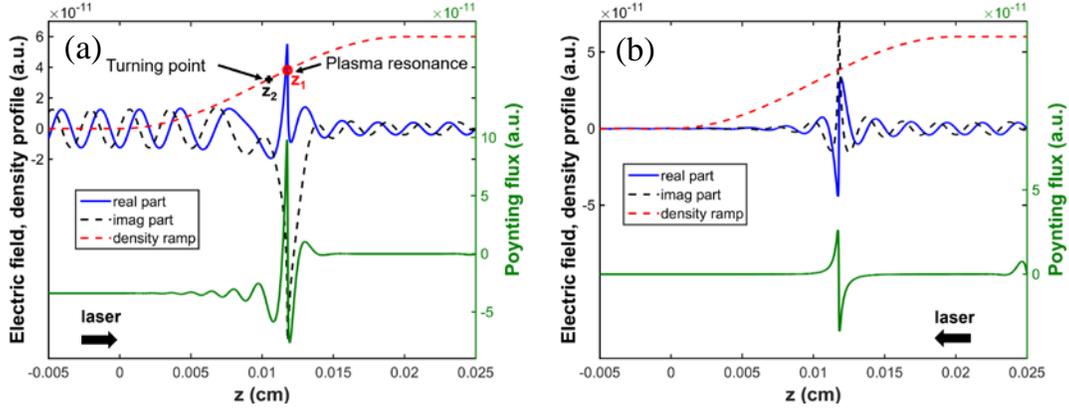

Figure 6: THz generation via resonant transition radiation in a diffuse (a) increasing and (b) decreasing density ramp. Shown are numerical result of the electric field described by Eq. (2) (real part, blue solid and imaginary part, black dashed), and the electron density profile (red dashed) with a 200 μm ramp length starting at $z = 0$ to $z = 200$ μm. The Poynting flux, described by Eq. (5), is also shown (green solid).

To better understand the mechanism of radiation generation we plot the Poynting flux, Eq. (5). It can be seen that only in the case of an increasing density ramp, Fig. 6a, is there an appreciable power flux leaving the plasma. The behavior of the Poynting flux in Fig. 6 shows that the THz radiation is generated at the plasma resonance $z_1$, where its frequency matches the local plasma frequency, i.e. the dielectric function $\varepsilon(z_1) = 0$. This is evidenced by the jump in Poynting flux that occurs near $z = z_1$. It is shown in Appendix B that when the ramp length is much greater than the THz wavelength, the particular and homogeneous solutions of Eq. (2) decouple except at the resonance, and the Poynting flux is carried by the homogeneous solution. To escape the plasma, the THz radiation must first tunnel to the turning point $z_2$, where the wavenumber $k(z_2) = 0$, before it leaves the plasma and propagates into the vacuum. The oscillations in Poynting flux seen for $z < z_1$ are due to the beating of the laser generated particular solution of Eq. (2) and the homogeneous solution. This picture will be used to develop a scaling formula subsequently.

As can be seen by comparing Figs 6a and 6b the radiation energy from an increasing ramp is greater than for a decreasing ramp. Further, comparison of the increasing and decreasing density cases shows that the power radiated in the increasing ramp exceeds that of the sharp boundary radiation, which in turn, exceeds the power radiated in the decreasing ramp. This agrees with the PIC simulations discussed in the context of Fig. 4. An analytical calculation confirming this phenomena is presented in detail in Appendix B.

We now compare the spectra of THz radiation predicted by the PIC simulations and by numerical solution of Eqs. (2-7). Figure 7 shows the energy per unit length and frequency ($dU'/d\omega$ in Eq. (4a)) versus frequency for several increasing ramps of varying length.

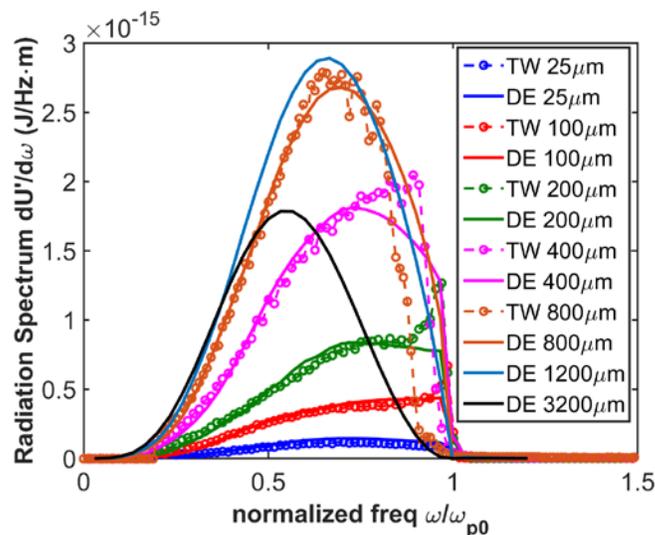

Figure 7: Comparisons of radiated spectral density between TurboWAVE and the proposed model for different increasing ramp lengths. Shown are TurboWAVE (TW) simulation results (dash, circle marker) and numerical results from the differential equation (DE) based on our developed model (solid) for different ramp lengths up to $3200\mu m$. Integration over frequency gives the total amount of radiated energy in 2D geometry; for example, the total energy is 0.1375 J/m for ramp length of $1200\mu m$, which is an enhancement of ~50 times compared with 0.0027 J/m for the sharp boundary.

The curves match closely except for a small interval of frequencies near the maximum plasma frequency. This difference could be due to the finite transverse size of the plasma in the PIC simulations or the slight difference in the density profiles. The PIC simulation has a fifth order polynomial representation of the density profile while the differential equation was solved for a sine-squared profile.

The dependence of the radiated flux, $S_z(\omega, k_x)$ defined by Eq. (5), on transverse wavenumber is shown in Fig. 8a for $\omega_1 = 0.6\omega_{p0}$, where $\omega_{p0}$ is the maximum plasma frequency. The flux is peaked at a wavenumber that decreases as the ramp length increases. This is shown in Fig. 8b where the wavenumber for the peak of the flux is plotted versus ramp length. The dependence of the peak wavenumber on scale length is a power law with a best-fit exponent $-0.3274$. For wavenumbers above the peak value the flux decreases rapidly, and the rate of decrease increases with ramp length.

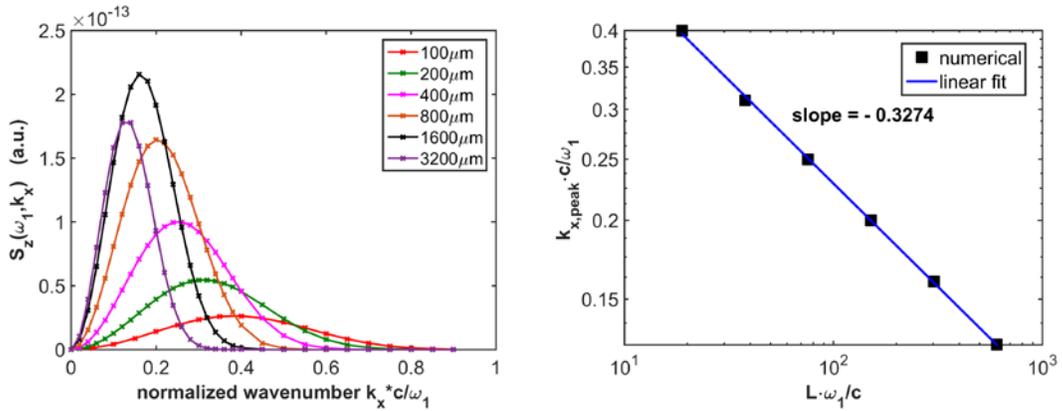

Figure 8: (a) Comparisons of $S_z$ (Eq. (5)) at a particular radiation frequency $\omega_1 = 0.6\omega_{p0}$ for different increasing ramp lengths. Shown are numerical results of our developed model (solid, cross marker) for different ramp lengths up to $3200\,\mu m$. (b) Peak radiation angle (corresponding $k_x$) agrees with our model $k_{x,peak} c / \omega \sim (c/L_0\omega)^{1/3} \ll 1$.

We now propose a simple scaling formula for the dependence of the radiated THz energy on ramp length based on the picture suggested by Fig. 6a. In our model, THz radiation is first generated by the laser pulse at the plasma resonance $\varepsilon(z_1) = 0$, and it must tunnel to the turning point $z_2$, where $\omega^2 \varepsilon(z_2) - k_x^2 c^2 = 0$. In the interval $z_2 < z < z_1$ the THz radiation has an imaginary wavenumber $k(z) = \sqrt{\omega^2 \varepsilon(z)/c^2 - k_x^2} = i\kappa$. If we assume that the density varies linearly with distance for $z_2 < z < z_1$, then $\varepsilon(z) = -(z - z_1 - iZ_\nu)/L_0$ and $z_2 = z_1 - L_0 k_x^2 c^2 / \omega^2$ where $L_0$ is the characteristic length defined by Eq. (B4) in Appendix B and $Z_\nu = L_0 \nu / \omega$. The THz radiation power escaping the plasma will then have a dependence on parameters due to the tunneling factor

$$S_z \propto \exp\left(-2\int_{z_2}^{z_1} \kappa(z) dz\right) = \exp\left(-\frac{4}{3} k_x^3 L_0 c^2 / \omega^2\right).$$

This dependence on transverse wavenumber will apply as long as the turning point $z_2$ and the plasma resonance $z_1$ are well separated. This in turn implies $k_x^2$ is larger than a critical value that is displayed in Fig. 8b.

To estimate the critical value of the transverse wavenumber we examine the differential equation, Eq. (2), for small $k_x^2 c^2 / \omega^2$ where the turning point $z_2$ and resonance $z_1$ are close together. We introduce a new axial coordinate, $\zeta = \mu(z - z_1 - iZ_\nu)$, where $\mu$ is a scale factor to be determined. We make the following replacement in Eq. (2), $\varepsilon = -\zeta (L_0 \mu)^{-1}$, $k^2 = -k_z^2 \zeta (L_0 \mu)^{-1} - k_x^2$ and $dz = d\zeta / \mu$, where $k_z = \omega / c$. This transforms Eq. (2) to the following:

$$\frac{d}{d\zeta}\left(\frac{\zeta}{\zeta + \zeta_k} \frac{d}{d\zeta} \overline{E}_x\right) - \frac{k_z^2}{\mu^3 L_0} \zeta \overline{E}_x = \frac{k_z^2}{\mu^2} S(z), \tag{8}$$

where $\zeta_k = k_x^2 L_0 \mu / k_z^2$ and the source $S(z)$ is given in Eq. (2).

We now pick a value of the scale factor $\mu$ so that both terms on the left of Eq. (8) are the same size, which leads to $\mu = (k_z^2 / L_0)^{1/3}$. Next, we notice that the turning point $\zeta = -\zeta_k$ and resonance $\zeta = 0$ will be close together when $\zeta_k = 1$ or $k_{x,peak} c / \omega \sim (c / L_0 \omega)^{1/3} \ll 1$. This determines the radiation angles (corresponding to $k_{x,peak}$) where the peak of THz radiation exists as shown in Fig. 8a, and is in agreement with the scaling of Fig. 8b.

We also estimate the size of the electric field in this case from Eq. (8), $\bar{E}_x \sim (k_z L_0)^{2/3} S$ and the source is evaluated as $S \sim -ik_x k_z^2 \mu^{-2} \hat{V}_p q_e^{-1} \exp(ik_z z)$, therefore we have the following expressions:

$$\bar{E}_x \sim -ik_z^2 L_0 \frac{\hat{V}_p}{q_e} \exp(ik_z z), \qquad (9a)$$

$$\frac{d\bar{E}_x}{dz} \sim i\mu \bar{E}_x \sim (k_z^4 L_0)^{2/3} \frac{\hat{V}_p}{q_e} \exp(ik_z z). \qquad (9b)$$

Using Eq. (9b), the peak value of $S_z$ as appears in Fig. 8a is estimated to be,

$$S_{z,peak} \sim \frac{c}{16\pi^3} k_z^2 (k_z L_0)^{5/3} \left(\frac{\hat{V}_p}{q_e}\right)^2. \qquad (10)$$

The radiated spectral density which is the integral of $S_z$ according to Eq. (4a) can be further estimated as the product $S_{z,peak} k_{x,peak}$ and scales as $L_0^{4/3}$. Based on this analysis one would expect the radiated spectral density $dU'/d\omega$ to increase monotonically with density scale length $L_0$. However, as observed in Figs. 5 and 7 the radiated spectral density reaches a maximum for $L_0 = 1200 \mu m$ and then decreases. This is explained by the presence of collisional damping. First there is local damping near $z = z_1$ where the THz is generated at the plasma resonance. In Appendix B, it is shown that the generated radiation is reduced by a factor $T_L = \exp(-2\omega Z_v / c) \sim \exp(-\alpha_L \nu L / c\pi)$ due to collisions. Another reduction due to damping occurs during propagation from the turning point to vacuum, i.e.,

$$T_P = \exp\left(2\int_{z_2}^{0} \left|\text{Im}\left(\sqrt{k^2}\right)\right| dz\right) \sim \exp(-\alpha_P \nu L / c\pi).$$

Therefore, the scaling formula for the radiated THz spectral density shown in Fig. 5 can be approximated as the following,

$$\frac{dU'}{d\omega} = \left(\frac{\omega}{c}\right)^3 \left(\frac{\hat{V}_p}{q_e}\right)^2 \left(\frac{\omega}{c}L_0\right)^{4/3} \exp\left[-(\alpha_L + \alpha_P)\frac{vL}{c\pi}\right], \quad (11)$$

where $\alpha_L$ and $\alpha_P$ both depend on radiation frequency $\omega$. Formula (11) matches well with the results based on solution of Eq. (2) as well as the PIC simulations shown in Fig. 5.

Finally, to further investigate the radiated energy in the decreasing ramp, we conducted PIC simulations using different ramp lengths. As predicted by our model, both PIC simulations and numerical solutions of Eq. (2) show that the amount of THz generated dramatically decreases in the case of a decreasing ramp. In fact, one can conclude that the THz generated in the decreasing ramp case is negligible (3 orders of magnitude less) compared with the amount of THz generated in the increasing density ramp case. The cause of this asymmetry is a form of phase matching as discussed in Appendix B.

## IV. CONCLUSIONS

We have both theoretically and numerically investigated ponderomotively driven resonant THz transition radiation generated at plasma boundaries. Broad-band THz radiation is generated with frequencies up to the maximum plasma frequency. The parameters of the driving pulse as well as the plasma profiles affect the properties of the generated THz radiation. The spectrum and angular distribution of the THz radiation can also be tuned by varying these parameters.

Resonant transition radiation is generated at a diffuse plasma boundary and is preferentially enhanced if the laser pulse propagates through an increasing density profile. We've developed a model to describe the physical processes in this diffuse plasma case. The THz is generated primarily at the plasma resonance and must tunnel to a turning point before it leaves the plasma and propagates into vacuum. The calculated Poynting flux shows that this process enhances the amount of THz energy efficiently through a density increasing ramp and diminishes THz generation through a decreasing density ramp. A scaling law was developed to allow one to estimate the amount of THz energy generated for different density ramp lengths.

Both numerical solutions of Eq. (2) and PIC simulations agree with our model. The amount of THz radiation generated can be dramatically increased compared with that in the sharp vacuum plasma boundary case[18]. As an example, a fixed driver pulse (1.66 J) excites approximately 422.9 µJ of THz radiation in a 1.2 mm increasing density ramp. Thus this mechanism provides the possibly of developing new high power tunable THz sources.

## ACKNOWLEDGEMENTS

The authors would like to acknowledge Dr. Daniel Gordon for the use of TurboWAVE. This work was supported by DOE under grant DESC0010741 and ONR/NRL under grant N00173131G018.

# APPENDIX A: 2D THEORY OF TRANSITION RADIATION

Following Ref. 18, we derive an expression for the radiated energy per unit frequency ($\omega$) and the radiation angle. We start from Maxwell's equations and the following cold plasma fluid equation,

$$m_e \frac{\partial \boldsymbol{J}}{\partial t} = q_e^2 n_e \boldsymbol{E} - q_e n_e \nabla V_p, \tag{A1}$$

where $m_e$, $q_e$ are the electron mass and charge, respectively. The quantity $n_e$ is the electron density, and $V_p$ is the averaged ponderomotive potential associated with the incident pulse.

As the pulse propagates through the vacuum plasma interface, the electric field of the low frequency radiation can be obtained from the following expression,

$$\frac{\partial^2 \boldsymbol{E}}{\partial t^2} + \omega_p^2 \boldsymbol{E} + c^2 \nabla \times (\nabla \times \boldsymbol{E}) = \frac{\omega_p^2}{q_e} \nabla V_p, \tag{A2}$$

where $\omega_p = (4\pi q^2 n_e / m_e)^{1/2}$ is the plasma frequency and $c$ is the speed of light in vacuum. An axisymmetric laser pulse with ponderomotive potential $V_p$ is described by Eq. (1).

To solve Eq. (A2) in 2D planar geometry, one can apply Fourier transform to Eq. (A2) with respect to both time $t$ and transverse coordinate $x$ according to the following definition:

$$\bar{E}(k_x, z, \omega) = \int_{-\infty}^{\infty} dt dx \boldsymbol{E}(x, z, t) \exp(i\omega t - i k_x x) \tag{A3a}$$

$$\boldsymbol{E}(x, z, t) = \frac{1}{(2\pi)^2} \int_{-\infty}^{\infty} d\omega dk_x \bar{E}(k_x, z, \omega) \exp(-i\omega t + i k_x x). \tag{A3b}$$

By simple calculations, one can obtain a differential equation as described by Eq. (2) for the Fourier transformed transverse component of the electric field $\bar{E}_x(k_x, z, \omega)$.

We now consider a sharp vacuum plasma interface such that the plasma density $n_e = 0$ for $z < 0$ (vacuum) and $n_e = n_0$ for $z \geq 0$ with an infinite transverse size. We solve Eq. (2) in the two uniform regions described above separately and note that the transverse component of electric and magnetic field are continuous at the boundary. We then inverse Fourier transform the field back to the space domain and obtain the following expression for the radiated electric field in vacuum ($z < 0$),

$$\bar{E}_x(x,z,\omega) = -i\sqrt{2\pi}\frac{\omega_p^2 R_L \tau_p}{8}\frac{m_e}{q_e}a_0^2\frac{\sin\chi\cos^2\chi}{\sqrt{i\omega r/c}}\cdot\frac{\exp\left(i\frac{\omega}{c}r - \frac{\omega^2\tau_p^2}{4} - \frac{\omega^2 R_L^2 \sin^2\chi}{4c^2}\right)}{\varepsilon\cos\chi + \sqrt{\varepsilon - \sin^2\chi}}, \quad (A4)$$

where $\chi$ is the observation angle in Fig. 1a, $r = \sqrt{x^2 + z^2}$ is the distance in 2D planar geometry and the dielectric function is given by $\varepsilon(\omega) = 1 - \omega_{p0}^2/\omega^2$ since $\nu \ll \omega$.

The radiated energy per unit length $U'$ is obtained using Eq. (4). One can also obtain the radiation spectrum in frequency and angle,

$$\frac{dU'}{d\omega} = \int_{-\infty}^{\infty} S_z(\omega, k_x) dk_x, \quad (A5a)$$

$$\frac{dU'}{d\chi} = \cos\chi \int_{-\infty}^{\infty} \frac{\omega}{c} S_z(\omega, k_x) d\omega, \quad (A5b)$$

where $k_x = \frac{\omega}{c}\sin\chi$.

By applying Eq. (A4) to Eq. (A5a), a formula for $dU'/d\omega$ describing the radiated energy per unit frequency ($\omega$) and length across the left diagnostic boundary in 2D planar geometry shown in Fig. 1a is obtained and displayed in Eq. (7).

## APPENDIX B: Analysis of Resonant Transition Radiation

In this appendix, we develop a model to describe THz generation in a density ramp via the process of resonant transition radiation. Note that the solution to Eq. (2) is a combination of a homogeneous solution (by simply setting the driver $S(z) = 0$) and a particular solution associate with the driver which describes the excitation of potential plasma wave fields. This particular solution in a uniform plasma can be analytically expressed as,

$$E_p = -\frac{ik_x}{\varepsilon}\frac{\omega_p^2}{\omega^2}\frac{\hat{V}_p}{q_e}\exp(ik_z z). \tag{B1}$$

The particular solution is valid when the density scale length is much longer than the THz wavelength, which is the case considered here. We note the particular solution breaks down at the plasma resonance, $\varepsilon(z_1) = 0$, where a more complete solution must be found.

It must be pointed out that this particular solution generates no radiation, which can be verified by substituting Eq. (B1) into Eq. (5). Therefore, the radiated energy is given by,

$$U' = \int_0^{+\infty} d\omega \int_{-\infty}^{+\infty} dk_x S_h(z), \tag{B2}$$

where the spectral density is given in terms of the homogeneous solution,

$$S_h(z) = \frac{c}{16\pi^3}\frac{1}{i}\frac{\omega}{c}\frac{\varepsilon}{k^2}\left(E_h^* \frac{dE_h}{dz} - E_h \frac{dE_h^*}{dz}\right), \tag{B3}$$

where $E_h$ is the solution from Eq. (2) by setting $S(z) = 0$. Equation (B3) also breaks down at the plasma resonance where the local particular and homogeneous solutions couple determining the amplitude of the homogeneous solution.

One can show that $S_h(z)$ is a constant (independent of $z$) for $z < z_1$, which is further verified by the numerical solutions shown in Fig. 6a. To determine the level of homogeneous solution we expand around the plasma resonance where $\varepsilon(z) \approx 0$. We take $z_1 = 0$ and the dielectric function can be approximated as $\varepsilon \approx -z/L_0 + i\nu/\omega$ where $L_0$ is the characteristic length defined as,

$$\frac{1}{L_0} = -\frac{d\operatorname{Re}\{\varepsilon\}}{dz}\Big|_{\operatorname{Re}\{\varepsilon\}=0}. \tag{B4}$$

Near the resonance, the homogeneous version of Eq. (2) becomes

$$\frac{d}{dz}\left(z\frac{d}{dz}E_h\right) = 0. \tag{B5}$$

And the solution of $E_h$ where $z \to 0^-$ ($z_1^-$) satisfies the following,

$$E_h = c_<\left(\ln\frac{z}{L_0} + \alpha_<\right). \tag{B6}$$

An expression of $S_h$ can be obtained by using the above approximation in (B3),

$$S_h = \frac{1}{16\pi^3}\frac{\omega}{ik_x^2 L_0}|c_<|^2\left(\alpha_<^* - \alpha_<\right), \tag{B7}$$

where $c_<$ and $\alpha_<$ are constants, which are determined by examining the solution of Eq. (2) near the plasma resonance including the source. A similar expression for the homogeneous solution for $z > 0$ applies but with different constants, $c_>$ and $\alpha_>$. We now derive an expression for the total solution near the resonance as shown following,

$$\bar{E}_x = \int_z \frac{\psi dz'}{(z'-iZ_\nu)}\exp(\pm ik_z z') + c\left[\ln\left(\frac{z}{L_0} - i\frac{\nu}{\omega}\right) + \alpha\right], \tag{B8}$$

where $\psi = k_x k_z L_0 \hat{V}_p / q_e$, $Z_\nu = L_0 \nu / \omega$, $c$, $\alpha$ are constants to be matched to the homogeneous solution and the integration contour is to be determined. The $\pm$ sign demotes the direction of the laser pulse. In Eq. (B8), the first term becomes the particular solution associated with the laser pulse far from the resonance and the second term is in the form of the homogeneous solution as described by Eq. (B6), which presumably will carry the radiation.

Expression (B7) shows that the Poynting flux is given by the imaginary part of the coefficients $\alpha_>$ and $\alpha_<$. For $z > 0$, the homogeneous solution is evanescent and consequently $\alpha_>$ is real. The integral expression (B8) relates the coefficients $\alpha_>$ and $\alpha_<$.

We now determine the connection between $\alpha_>$ and $\alpha_<$. If the laser is traveling in the positive-z direction ($+ik_z$), the upper limit of the limit should be taken to be $z' = i\infty$ to give convergence to the integral. The integration contour is the then of the type shown in the top row of Fig. B1. Then for $z > 0$ the path is $C$ as illustrated on the top left of Fig. B1. Evaluation of Eq. (B8) using contour $C$, one can fix $c$ and $\alpha$ in (B8) to be $c_>$ and $\alpha_>$.

$$\bar{E}_x = \int_z^{i\infty} \frac{\psi dz'}{(z'-iZ_v)} \exp(ik_z z') + c_> \left[ \ln\left(\frac{z}{L_0} - i\frac{v}{\omega}\right) + \alpha_> \right]. \tag{B9a}$$

For $z < 0$, the contour is deformed to the one labeled $C'$ and a residue at the pole $z' = iZ_v$ is accumulated.

$$\bar{E}_x = \int_z^{i\infty} \frac{\psi dz'}{(z'-iZ_v)} \exp(ik_z z') + \left(2\pi\psi \exp(-k_z Z_v) + c_> \left[ \ln\left(\frac{z}{L_0} - i\frac{v}{\omega}\right) + \alpha_> \right]\right). \tag{B9b}$$

Matching Eq. (B6) for $z > 0$ and $z < 0$ to Eqs. (B9a) and (B9b) gives

$$c_< \left[ \ln\left(\frac{z}{L_0} - i\frac{v}{\omega}\right) + \alpha_< \right] = 2\pi\psi \exp(-k_z Z_v) + c_> \left[ \ln\left(\frac{z}{L_0} - i\frac{v}{\omega}\right) + \alpha_> \right]. \tag{B10}$$

Matching the logarithmic and constant parts, the following conditions are obtained,

$$c_< = c_> = \frac{2\pi\psi \exp(-k_z Z_v)}{\alpha_< - \alpha_>}. \tag{B11}$$

By substituting Eq. (B11) into Eq. (B7), finally we obtain an expression for the total radiated energy per unit frequency $\omega$ and per unit wavenumber $k_x$ when the laser pulse is incident on an increasing density ramp,

$$S_h = \frac{1}{4\pi} \frac{\omega}{ik_x^2 L_0} \frac{(\alpha_<^* - \alpha_<)}{|\alpha_< - \alpha_>|^2} |\psi \exp(-k_z Z_v)|^2, \tag{B12}$$

where $\alpha_<$ and $\alpha_>$ are constants that must be determined by numerical solution of the homogeneous equation.

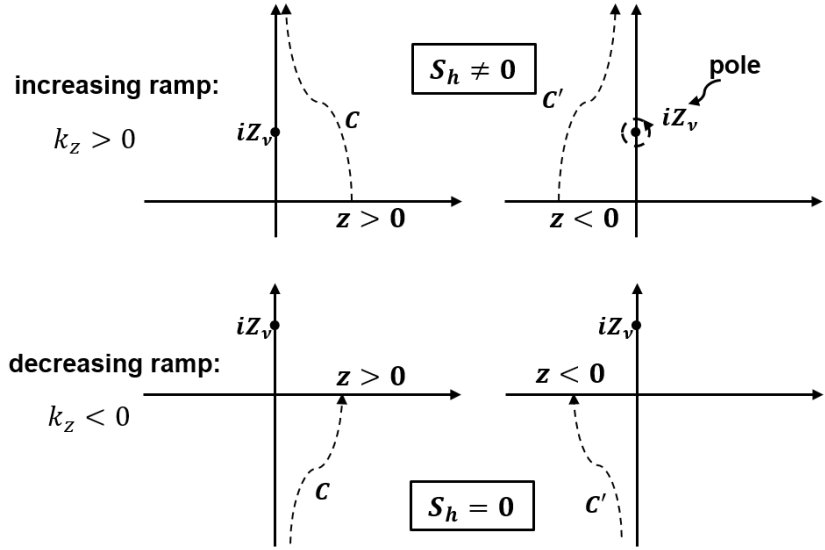

Figure B1: Contours for evaluation of the integral in Eq. (B8) for an increasing (a) and decreasing (b) density ramp.

If the pulse travels from right to left ($k_z < 0$), it propagates through a decreasing density ramp. As a result of $k_z < 0$, the integral in Eq. (B8) must be carried out on a contour that terminates in the lower half plane as illustrated in Fig. B1b. The solutions for $z > 0$ and $z < 0$ in this case have no jump since there is no pole (singularity) on the negative image axis (Fig. B1). One finds from Eq. (B10) without the residue contribution $c_> = c_<$ and $\alpha_> = \alpha_<$. Thus $S_h = 0$, meaning that no radiation is generated through the decreasing density ramp. This verifies our simulation results with TurboWAVE[20] in Fig. 4, where one can see the amount of THz generation through an increasing ramp is far more than that through the decreasing ramp. The physics of this asymmetry between the increasing and decreasing ramps can be explained by a form of phase matching[21]. The phase of the integrand in Eq. (B8) is given by $\phi + k_z z'$ where $\phi = \tan^{-1}(Z_v / z')$ decreases as $z'$ increases. Therefore, if $k_z > 0$, the phase matching condition can be satisfied at a point $z'$ where $k_z + d\phi/dz' = 0$. However, the phase matching condition cannot be satisfied if $k_z < 0$.

## APPENDIX C: NUMERICAL METHOD FOR EQUATION (2)

In Sec. III, we presented numerical results of solutions for the electric field $\bar{E}_x$ determined by Eq. (2). Here we present details of the method. Eq. (2) can be put in the general form,

$$\frac{d}{dz}\left(p(z)\frac{d}{dz}\bar{E}_x\right) + q(z)\bar{E}_x = S(z), \tag{C1}$$

where $p(z) = \frac{\varepsilon(z)}{k^2(z)}$, $q(z) = \varepsilon(z) = 1 - \frac{\omega_p^2(z)}{\omega(\omega+i\nu)}$ and $k^2(z) = \frac{\omega^2}{c^2}\varepsilon(z) - k_x^2$.

We take $z = 0$ to be the boundary between the plasma and vacuum. One boundary condition is that the field $\bar{E}_x$ should be an outgoing wave in vacuum ($z < 0$). Thus for $z < 0$, in vacuum,

$$\bar{E}_x = C_1 \exp(-ik_1 z). \tag{C2a}$$

We assume the density is uniform for $z > L$, then

$$\bar{E}_x = C_2 \exp(-\kappa z) + E_p(z), \tag{C2b}$$

where $C_1$, $C_2$ are two constants to be determined and $k_1^2 = \frac{\omega^2}{c^2} - k_x^2$ and $\kappa^2 = k_x^2 - \frac{1}{c^2}\left(\omega^2 - \frac{\omega_{p0}^2}{1+i\nu/\omega}\right)$. The function $E_p$ is the particular solution described by Eq. (B1).

The homogeneous solutions are exponentially growing and decaying with $z$ for $z > z_2$ (see Fig. 6). Thus, direct integration of Eq. (C1) for large ramp lengths is subject to errors. To overcome this, we then use the method of variation of parameters by first considering two homogeneous solutions $u_1$ and $u_2$ of Eq. (C1) (taking the source term $S(z) = 0$), which satisfy the following boundary conditions,

$u_1(z)$:  $u_1(z) \to 0$ as $z \to +\infty$, and

$u_2(z)$: $\quad u_2(z) \to$ outgoing wave as $z \to -\infty$.

We then write the general solution of Eq. (C1) in the following form,

$$\bar{E}_x(z) = A_1(z)u_1(z) + A_2(z)u_2(z). \tag{C3}$$

Since we have introduced two functions, $A_1$ and $A_2$, to represent a single unknown, we are free to make up a second relation between them,

$$u_1 \frac{dA_1}{dz} + u_2 \frac{dA_2}{dz} = 0. \tag{C4}$$

Substituting Eq. (C3) into Eq. (C1) and applying the condition (C4), one arrives at equations for the two functions, $A_1$ and $A_2$,

$$\frac{dA_1}{dz} = -\frac{u_2 S}{W}, \tag{C5a}$$

$$\frac{dA_2}{dz} = \frac{u_1 S}{W}. \tag{C5b}$$

The quantity $W$ is the Wronskian,

$$W = p\left(u_1 \frac{du_2}{dz} - u_2 \frac{du_1}{dz}\right), \tag{C6}$$

and $W$ is independent of $z$. As an example, shown in Fig. C1 is the Wronskian formed from numerical solutions for $u_1$ and $u_2$ for a 200 μm increasing density ramp. Other parameters are the same as in Sec. III. The relative variation is within 0.005%.

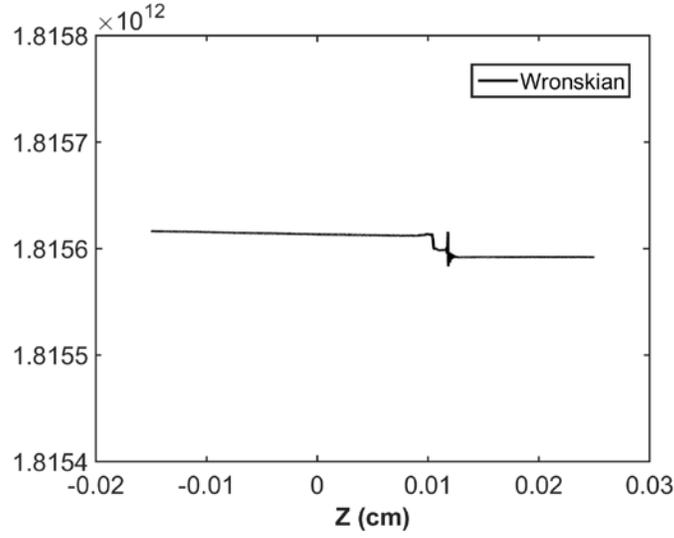

Figure C1: Wronskian described by Eq. (C6) for a $200\,\mu\text{m}$ increasing density ramp.

After obtaining the Wronskian from the simulation, we can further numerically solve for the two coefficient functions, $A_1$ and $A_2$, and finally obtain the generated THz field $\bar{E}_x$ in vacuum, $z<0$. In Sec. III, Fig 6 is an example of the simulation result for a particular radiation frequency and transverse wavenumber. Scanning over all possible radiation frequencies and wavenumbers is conducted to obtain the total radiated THz energy into vacuum using Eq. (B2).